\documentclass{article}
\usepackage{spconf}
\usepackage{amsmath,amsfonts,amsthm,amssymb}
\usepackage{graphicx,float,wrapfig}
\usepackage{graphicx}
\usepackage{graphics,color}
\usepackage{mathtools}
\usepackage{nicefrac}
\usepackage[T1]{fontenc}
\usepackage{upquote}
\usepackage{algorithm,algorithmic}
\usepackage{caption}% http://ctan.org/pkg/caption
\makeatletter
\@addtoreset{algorithm}{section}% algorithm counter resets every chapter
\makeatother
\newtheorem{theorem}{Theorem}[section]
\newtheorem{definition}[theorem]{Definition}

\numberwithin{equation}{section}

\def\x{{\mathbf x}}

\title{Off-The-Grid Spectral Compressed Sensing With Prior Information}
\name{Kumar Vijay Mishra, Myung Cho, Anton Kruger and Weiyu Xu}
\address{Department of Electrical and Computer Engineering\\The University of Iowa, Iowa City IA 52242}

\begin{document}
\ninept
\maketitle
\begin{abstract}
Recent research in off-the-grid compressed sensing (CS) has demonstrated that, under certain conditions, one can successfully recover a spectrally sparse signal from a few time-domain samples even though the dictionary is continuous. In this paper, we extend off-the-grid CS to applications where some prior information about spectrally sparse signal is known. We specifically consider cases where a few contributing frequencies or poles, but not their amplitudes or phases, are known \emph{a priori}. Our results show that equipping off-the-grid CS with the known-poles algorithm can increase the probability of recovering all the frequency components.
\end{abstract}
\begin{keywords}
compressed sensing, spectral estimation, basis mismatch, matrix completion, known poles
\end{keywords}
\section{Introduction}
\label{sec:intro}
Compressed sensing (CS) is a new sampling paradigm which postulates uniting two critical steps involved in processing a signal: digital data acquisition and its compression \cite{donoho2006compressed} \cite{kutyniok2012theory}. To recover a signal from fewer random measurements, CS algorithms harness the inherent sparsity of the signal under some appropriate basis or dictionary.

For example, consider a frequency-sparse signal $x[l]$ represented as a sum of $s$ complex exponentials,
\begin{equation}
\label{eq:sigmodelstd}
x[l] = \sum\limits_{j=1}^{s} c_je^{i2\pi f_jl} = \sum\limits_{j=1}^{s} |c_j|a(f_j, \phi_j)[l]\phantom{1}, \phantom{1} l \in \mathcal{N}
\end{equation}
where $c_j = |c_j|e^{i\phi_j}$ ($i = \sqrt{-1}$) represents the complex coefficient of the frequency $f_j \in [0, 1]$, with amplitude $|c_j| > 0$, phase $\phi_j \in [0, 2\pi)$, and frequency-\textit{atom} $a(f_j, \phi_j)[l] = e^{i(2\pi f_j l + \phi_j)}$. We use the index set $\mathcal{N} = \{l\phantom{1}|\phantom{1} 0 \le l \le n-1\}$, where $|\mathcal{N}| = n, n \in \mathbb{N}$, to represent the time samples of the signal. It is customary to label only the frequency information - either the exponentials $e^{i2\pi f_jl}$ or just $f_j$ - as $poles$ \cite{majda1989simple} \cite{wei1990new}.

When $f_j$ takes values only on a \textit{discrete frequency grid}, the Discrete Fourier Transform (DFT) matrix can be used as an appropriate finite discrete dictionary for the sparse representation of $x[l]$. However, it is quite possible for the true frequencies to be anywhere in the \textit{continuous domain} $[0, 1]$. Since the true continuous-domain frequencies may lie off the center of the DFT bins, the DFT representation in this case would destroy the sparsity of the signal and result in the so-called \textquotedblleft basis mismatch\textquotedblright\phantom{1}\cite{chi2011sensitivity}. This can be mitigated to a certain extent by finer discretization of the DFT grid. But that could lead to higher correlation of the sensing matrix and, thus, computationally infeasible or expensive signal recovery \cite{duarte2013spectral}.

These numerical problems associated with the spectral spill-over in the Dirichlet kernel have recently been addressed by the \textit{off-the-grid compressed sensing} approach \cite{candes2013towards} \cite{tang2012csotg}. This method relies on atomic norm minimization and guarantees recovery of frequencies lying anywhere in the continuous domain [0, 1] from a limited number of random observations, provided the line spectrum satisfies nominal resolution conditions. A two-dimensional generalization to this method involves Hankel matrix completion and guarantees robustness against corruption of data \cite{chi2013robust}.

These approaches to off-the-grid frequency recovery assume that, other than the sparse nature of signal frequencies, little is known about the signal \textit{a priori}. However, in many applications such as radar \cite{skolnik2008radar}, acoustics \cite{trivett1981modified}, and power systems \cite{zygarlicki2012prony}, some information about the signal may already be known through previous measurements or known electrical properties of the signal source. For example, a radar engineer may be aware of the Doppler signature of standard targets such as a fighter jet or a commercial airplane \cite{skolnik2008radar}. Similarly, in a weather radar scenario, the range of Doppler frequencies and the spectrum widths for certain types of precipitation events (storms or tornadoes) are often known from previous observations \cite{doviak1993doppler}.

In this work, we consider a frequency-sparse signal composed of multiple frequencies, of which a few frequencies are known \emph{a priori}. However, the amplitudes and phases of these frequencies are not known. We then propose the \textit{known-poles} algorithm based on the \textit{conditional} atomic norm minimization to first recover the complex coefficients of these known frequencies. Once the known frequencies are filtered from the original signal, the problem then simplifies to a search for the remaining frequency components in the signal.
\subsection{Main Results}
\label{ssec:intro_results}
Suppose $p$ out of $s$ frequencies contributing to $x[l]$ are known \textit{a priori}. Suppose that $f_j$'s, $s-p+1\leq j\leq s$, are known frequencies. Then, \begin{equation}
\label{eq:sigmodelprior}
x[l] = \sum\limits_{j=1}^{s-p} c_je^{i2\pi f_j l} + \sum\limits_{j=s-p+1}^{s} d_je^{i2\pi f_{j}l}, \phantom{1} l \in \mathcal{N}
\end{equation}
where $c_j$ and $d_j$ are \emph{unknown} complex coefficients of these unknown and known frequencies respectively. Our main result in this work demonstrates that if the known frequency information is included in the off-the-grid atomic norm minimization approach, then all the remaining frequencies can be exactly recovered with a higher probability, even though all the frequencies are \textit{continuous-domain} frequencies. Further, the recovery of the entire unknown spectral content using \textit{known-poles} algorithm is possible with a smaller number of random observations. When the frequencies do not satisfy any minimum resolution conditions, our algorithm suffers lesser degradation in the spectral recovery performance compared to the algorithm which does not use any prior information.
\subsection{Relation To Prior Work}
\label{ssec:intro_prior}
Estimation of the parameters of a frequency-sparse signal has been a long-pursued problem\cite{prony1795}. A description of several common tools and methods dealing with the line spectral estimation of regularly sampled signals in the presence or absence of noise can be found in \cite{marple1987digital} and \cite{stoica2005spectral}. These methods have often been accompanied by an undercurrent of research on the spectral analysis with available prior information. For example, the classical Prony's method can be easily modified to account for known frequencies \cite{trivett1981modified}. Variants of the subspace-based frequency estimation methods such as MUSIC and ESPRIT have also been formulated \cite{linebarger1995incorporating} \cite{wirfalt2011subspace}, where prior knowledge can be incorporated for parameter estimation. For applications wherein only approximate knowledge of the frequencies is available, the spectral estimation described in \cite{zachariah2013line} applies circular von Mises probability distribution on the spectrum.

For irregularly spaced samples, sparse signal recovery methods which leverage on prior information have recently gained attention for general applications \cite{ji2008bayesian} \cite{amin2009weighted} \cite{vaswani2010modified} as well as, specifically, spectral estimation \cite{bourguignon2007sparsity}. To the best knowledge of the authors, the CS-based spectral estimation methods - aforementioned and others - have not considered spectral components lying off the given frequency grid.

Our model and methods for recovery of the frequency-sparse signal in the continuous dictionary are inspired by \cite{tang2012csotg} and \cite{candes2013towards}. We additionally show that, if some poles are known, the performance of the recovery can be improved in terms of the probability of successful recovery and the number of random samples needed.
\section{System Model}
\label{sec:sysmodel}
The signal in (\ref{eq:sigmodelstd}) can be modeled as a positive linear combination of the unit-norm frequency-\textit{atoms} $a(f_j, \phi_j)[l] \in \mathcal{A} \subset \mathbb{C}^n$ where $\mathcal{A}$ is the set of all frequency-atoms.  These frequency atoms are basic units for synthesizing the frequency-sparse signal. Further, suppose the signal in (\ref{eq:sigmodelstd}) is observed on the index set $\mathcal{M} \subset \mathcal{N}$, $|\mathcal{M}| = m \ll n$ where $m$ observations are chosen uniformly at random. Then, to estimate the remaining $\mathcal{N} \setminus \mathcal{M}$ samples of the signal $x$, \cite{chandrasekaran2012theconvex} suggests minimizing the atomic norm $||\hat{x}||_\mathcal{A}$ - a sparsity-enforcing analog of $\ell_1$ norm for a general atomic set $\mathcal{A}$-among all vectors $\hat{x}$ leading to the same observed samples as $\x$. The atomic norm is given by,
\begin{equation}
\label{eq:atomicnorm}
||\hat{x}||_{\mathcal{A}} = \underset{c_j, f_j}{\text{inf}}\phantom{1}\left\{\sum\limits_{j=1}^s|c_j|: \hat{x}[l] = \sum\limits_{j=1}^{s} c_je^{i2\pi f_jl} \phantom{1}, \phantom{1} l \in \mathcal{M}\right\}
\end{equation}
The semidefinite formulation of $||\hat{x}||_\mathcal{A}$ is defined as follows:
\begin{definition} \cite{tang2012csotg} Let $T_n$ denote the $n \times n$ positive semidefinite Toeplitz matrix, $t \in \mathbb{R}^+$, Tr($\cdot$) denote the trace operator and $(\cdot)^*$  denote the complex conjugate. Then,
    \begin{equation}
	||\hat{x}||_{\mathcal{A}} = \underset{T_n, t}{\text{inf}} \left\{\dfrac{1}{2|\mathcal{N}|} \text{Tr($T_n$)} + \frac{1}{2}t : \begin{bmatrix*}[r] T_n & \hat{x} \\ \hat{x}^* & t \end{bmatrix*} \succeq 0 \right\}
	\end{equation}
\end{definition}
The positive semidefinite Toeplitz matrix $T_n$ is related to the frequency atoms through the following result by Carath{\`e}odory \cite{cara1911uber}:
\begin{align}
T_n &= URU^*\\
\text{where } U_{l'j'} &= a(f_{j'}, \phi_{j'})[l'],\\
                    R &= \text{diag}([b_1, \cdots, b_{r'}])
\end{align}
The diagonal elements of $R$ are real and positive, and $r' = \text{rank}(T_n)$.

Consistent with this definition, the atomic norm minimization problem for the frequency-sparse signal recovery can now be formulated in a semidefinite program (SDP) with $m$ affine equality constraints:
\begin{flalign}
	\label{eq:semiotg}
	& \underset{T_n, \hat{x}, t}{\text{minimize}}\phantom{1} \dfrac{1}{2|\mathcal{N}|} \text{Tr($T_n$)} + \frac{1}{2}t\nonumber\\
	& \text{subject to}\phantom{1} \begin{bmatrix*}[r] T_n & \hat{x} \\ \hat{x}^* & t \end{bmatrix*} \succeq 0\\
	& \hat{x}[l] = x[l], \phantom{1} l \in \mathcal{M}\nonumber
\end{flalign}
		
\section{Known-Poles Algorithm}
\label{sec:algoknown}
We now consider the case when some frequency components are known \emph{a priori} but their corresponding amplitudes and phases are not. A common approach to harness the prior information about the sparse signal is to replace the classical $\ell_1$ norm with the weighted $\ell_1$ norm \cite{amin2009weighted} \cite{vaswani2010modified}. However, the \textit{known poles} signal does not lead to a trivial application of the weighted $\ell_1$ approach, since all the frequencies have a continuous domain. Therefore, we propose the \textit{conditional} atomic norm minimization as the \textit{known-poles} algorithm.

Let the index set of all the frequencies be $\mathcal{S}$, $|\mathcal{S}| = s$. Let $\mathcal{P}$ be the index set of all the known frequencies, and $|\mathcal{P}| = p$. Namely, we assume that the signal $x$ contains some known frequencies $f_j$, $j \in \mathcal{P} \subseteq \mathcal{S}$, $|\mathcal{P}| = p$. For unknown frequencies, let us denote their \emph{complex} coefficients as $d_j$ and their \textit{phaseless frequency atoms} as $\alpha_{j}[l] = a(f_j, 0)[l]  = e^{i2\pi f_j l}$. We define the \textit{conditional atomic norm} $||\hat{x}||_{\mathcal{A}|\mathcal{P}}$ for the \textit{known poles} as follows:\\
\begin{flalign}
    \label{eq:condatomicnorm}
    & ||\hat{x}||_{\mathcal{A}|\mathcal{P}} = \underset{c_j, d_j, f_j}{\text{inf}}\phantom{1}\left\{\sum\limits_{j=1}^{s-p}|c_j|: \hat{x}[l] = \sum\limits_{j=1}^{s-p} c_je^{i2\pi f_jl} \right.\nonumber\\
    & \phantom{1}\phantom{1}\phantom{1}\phantom{1}\phantom{1}\phantom{1}\phantom{1}\phantom{1}\phantom{1}\phantom{1}\phantom{1}\phantom{1}\phantom{1}\phantom{1}\phantom{1}\phantom{1}\phantom{1}\left.+ \sum\limits_{j=s-p+1}^{s} d_je^{i2\pi f_jl}\phantom{1}, \phantom{1} l \in \mathcal{M}\right\}
\end{flalign}
The semidefinite formulation for $||\hat{x}||_{\mathcal{A}|\mathcal{P}}$ is given as follows (proof omitted due to space limitations):
\begin{definition}
The conditional atomic norm for a vector $\hat{x}$ is given by
	\begin{equation}
	||\hat{x}||_{\mathcal{A}|\mathcal{P}} = \underset{T_n, \tilde{x}, t, d_j}{\text{inf}}\left\{\dfrac{1}{2|\mathcal{N}|} \text{Tr($T_n$)} + \frac{1}{2}t : \begin{bmatrix*}[r] T_n & \tilde{x} \\ \tilde{x}^* & t \end{bmatrix*} \succeq 0 \right\}
	\end{equation}
where $\tilde{x}[l] = \hat{x}[l] - \sum\limits_{j \in \mathcal{P}}\alpha_j[l]d_j$ represents the positive combination of complex sinusoids with unknown poles.
\end{definition}
The conditional atomic norm minimization problem can be posed as the following semidefinite formulation in a similar way as in (\ref{eq:semiotg}):
	\begin{flalign}
	\label{eq:semiprior}
	& \underset{T_n, \hat{x}, \tilde{x}, t, d_j}{\text{minimize}}\phantom{1} \dfrac{1}{2|\mathcal{N}|}  \text{Tr($T_n$)} + \frac{1}{2}t\nonumber\\
	& \text{subject to}\phantom{1} \begin{bmatrix*}[r] T_n & \tilde{x} \\ \tilde{x}^* & t \end{bmatrix*} \succeq 0\\
	& \hat{x}[l] = x[l], \phantom{1} l \in \mathcal{M}\nonumber\\
	& \hat{x}[l] = \tilde{x}[l] + \sum\limits_{j \in \mathcal{P}}\alpha_j[l]d_j, \phantom{1} l \in \mathcal{M}\nonumber
	\end{flalign}
$\tilde{x}$ can be viewed as the signal filtered of the \textit{known poles}. The remaining unknown frequencies in $\tilde{x}$ can be identified by the frequency localization approach \cite{tang2012csotg} based on computing the dual polynomial, which we restate for $\tilde{x}$ in Algorithm \ref{alg:freqLocal}.
\begin{algorithm}
\caption{\textit{Known-poles} algorithm}
\label{alg:freqLocal}
    \begin{algorithmic}[1]
    \scriptsize
    \STATE Solve the semidefinite program (\ref{eq:semiprior}) for \textit{conditional} atomic norm minimization to obtain $\tilde{x}$.
    \STATE Solve the following dual problem to obtain the optimum solution $q^{\star}$
	\begin{flalign}
		& \underset{q}{\text{maximize}}\phantom{1} \langle q, \tilde{x}\rangle_{\mathbb{R}}\nonumber\\
		& \text{subject to}\phantom{1} ||q||^*_\mathcal{A} \le 1\\
		& q[l] = 0, \phantom{1} l \in \mathcal{N} \setminus \mathcal{M}\nonumber
	\end{flalign}
    where $||\cdot||^*$ denotes the dual-norm.
	\STATE The unknown frequencies $f_j$, identify as $\left|\langle q^{\star}, \alpha_j \rangle\right| = 1$ where $j \in \mathcal{P}$. For $j \notin \mathcal{S} \setminus \mathcal{P}$, $\left|\langle q^{\star}, \alpha_j \rangle\right| < 1$.
	\STATE All the unknown frequencies identified, hereafter their corresponding complex coefficients can be recovered by solving a system of simultaneous linear equations $\tilde{x}[l] - \sum\limits_{j \in \mathcal{S} \setminus \mathcal{P}}c_j\alpha_j[l] = 0$.
    \end{algorithmic}
\end{algorithm}
\section{Performance Analysis}
\label{sec:perfanal}
In this section, we set out to give necessary and sufficient conditions under which (\ref{eq:semiprior}) recovers existing frequencies $f_j$, $1\leq j \leq s$ and their complex coefficients, regardless of what amplitudes these $s$ frequency components take. We first need to define a notion of ``feasible matrix'' $A \in \mathbb{C}^{m \times (n+s)}$. Recall that $f_r,\; 1 \leq r \leq s,$ are existing frequencies, and  $f_r,\; s-p+1 \leq r \leq s,$ are known frequencies. For a matrix $A \in \mathbb{C}^{m \times (n+s)}$ and an index set $K \subseteq \{1,2,...,n+s\}$, we define $\mathcal{D}(A_K)$ as the dictionary (set) of columns in $A$ corresponding to the index set $K$. We further define a few index sets as follows:
\begin{align*}
    & \mathcal{S}^{*} = \{r| n+1 \leq r \leq n+s \} \\
    & \mathcal{P}^{*} = \{r| n+s-p+1 \leq r \leq n+s \} \\
    & \mathcal{N}^{*} = \{r| 1 \leq r \leq n \} \\
    & \mathcal{N^{*}} \cup (\mathcal{S^{*} \setminus P^{*}})  = \{r| 1 \leq r \leq n+s-p \}
\end{align*}
In the following definition of a  ``feasible matrix'' $A \in \mathbb{C}^{m \times (n+s)}$, we will use $A_{\mathcal{S}^{*}}$ to accommodate the basis vectors for existing frequencies, and $A_{\mathcal{P}^{*}}$ to accommodate the basis vectors for known frequencies.
\begin{definition}
A matrix $A \in \mathbb{C}^{m \times (n+s)}$ is called a \emph{feasible} matrix if
 \begin{itemize}
 \item ${A}_r = a( f_{r-n}, \phi_{r-n})[{\mathcal{M}}],\; n+1 \leq r \leq n+p+s$,
where $\mathcal{M}$ is the sampling time index set;
\item $A_r = a(\widetilde{f_r}, \widetilde{\phi_r})[{\mathcal{M}}],\; 1\leq r \leq n$, where $\widetilde{f_r},\; 1 \leq r \leq n$, are $n$ distinct frequencies in [0,1]; and $\widetilde{\phi_r} \in [0,2\pi),\; 1 \leq r \leq n$;
\item $f_j \notin \{ \widetilde{f_r}| 1 \leq r \leq n\}$ when $s-p+1 \leq j \leq s$.
\end{itemize}
\end{definition}
We use $\textbf{N}(A)$ to denote the null space of $A$. We are now ready to give a statement of the conditions for successful recovery.
\begin{theorem}
Let $\displaystyle x[l] = \sum_{j=1}^{s}c_j e^{i 2\pi f_jl}=\sum_{j=1}^{s} |c_j|a(f_j, \phi_j)[l]$, $l \geq 0$. Assume that $\{f_j~|~s-p+1\leq j \leq s \}$ are $p$ known frequencies (but their amplitudes and phases are still unknown).  Then, for all possible amplitudes $|c_j|>0$, $1\leq j \leq s$,  the \emph{known-poles algorithm} uniquely recovers all the frequencies, their corresponding phases and amplitudes, if and only if the following holds.

For every feasible matrix $A\in \mathbb{C}^{m \times (n+s)}$ and for every $h \in \{~h~|~h \in \textbf{N}(A),\; h \neq 0,\; h_{ \mathcal{N^{*}} \cup (\mathcal{S^{*} \setminus P^{*}}) } \in \mathbb{R}^{n+s-p}, \; h_{\mathcal{S^{*}\setminus P^{*}}} <0,\;  h_{\mathcal{N^*}} \geq 0 \}$, $|| h_{ \mathcal{N^{*}}} ||_1 > || h_{\mathcal{S^{*}\setminus P^{*}}} ||_1 $ unless the following three statements hold true simultaneously.
\begin{align*}
    & (1)\;  a(f_j, \phi_j) \in \mathcal{D}(A_{\mathcal{N^*}}), when~1\leq j\leq s-p; \\
    & (2)\; h_{r_1} = - h_{r_2}, \\
    &whenever \;r_1 \in \mathcal{N^{*}},~r_2\in \mathcal{S^{*}\setminus P^{*}},\; and \; A_{r_2} = A_{r_1}; \\
    & (3)\; h_{r_3} = 0,\\
    & \quad \;if \; r_3 \in \mathcal{\mathcal{N}^*\cup \mathcal{S}^*}\; and \; A_{r_3} \notin \mathcal{D}(A_{S^{*}\setminus P^{*}}).
\end{align*}
\label{thm:iffcondition}
\end{theorem}
Due to space limitations, we omit the proof in this paper. We remark that, intuitively, when the size of known frequencies (namely $|\mathcal{P^*}|$) increases,  it may be easier to satisfy the key condition $\| h_{ \mathcal{N^{*}}} \|_1 > \| h_{\mathcal{S^{*}\setminus P^{*}}} \|_1 $ in Theorem \ref{thm:iffcondition} since, roughly speaking, $\| h_{\mathcal{S^{*}\setminus P^{*}}} \|_1$ decreases when $|\mathcal{P^*}|$ increases. So we expect that when more prior information is known, it becomes easier to recover all the frequencies.
\section{Numerical Simulations}
\label{sec:numsim}
We evaluated the \textit{known-poles} algorithm through a number of simulations using SDPT3 \cite{tutuncu2003solving} to solve the semidefinite program.  In all our experiments, the $s$ frequencies of the artificially generated signal were drawn at random in the band $[0, 1]$. Except for Experiment 4, the sampled frequencies were also constrained to have the minimum modulo spacing of $\Delta f = \nicefrac{1}{\lfloor (n-1)/4 \rfloor}$ between the adjacent frequencies. This is the theoretical resolution condition for the results in \cite{tang2012csotg}, although numerical experiments suggested that frequencies could be closer, i.e., $\Delta f$ could be $\nicefrac{1}{(n-1)}$. While working with the \textit{known poles}, we draw the first known frequency uniformly at random from the set of $s$ frequencies. As the number $p$ of \textit{known poles} increases, we retain the previously drawn known frequencies and draw the next known frequency uniformly at random from the remaining set of existing signal frequencies.

\begin{figure} \centering
\includegraphics[width=0.45\textwidth]{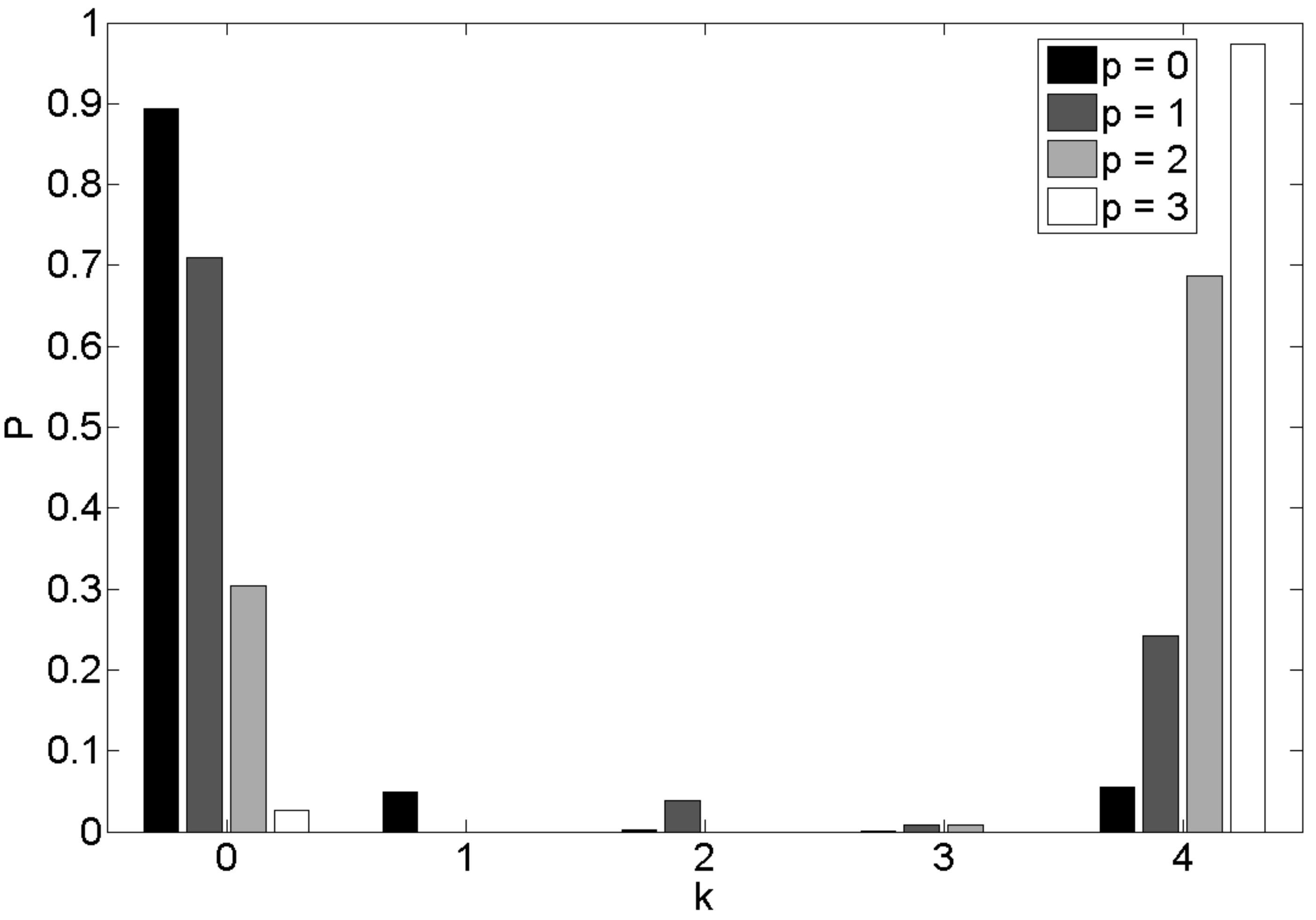}
\caption{The probability $P$ of recovering the unknown spectral content. The probability is computed for 1000 random realizations of the signal for the triple $(n, m, s) = (32, 9, 4)$. (For $k > 0$, $k \le p$ being the invalid cases, the corresponding bars have been omitted.)}
\label{fig:lowdim}
\end{figure}

\begin{figure} \centering
\includegraphics[width=0.45\textwidth]{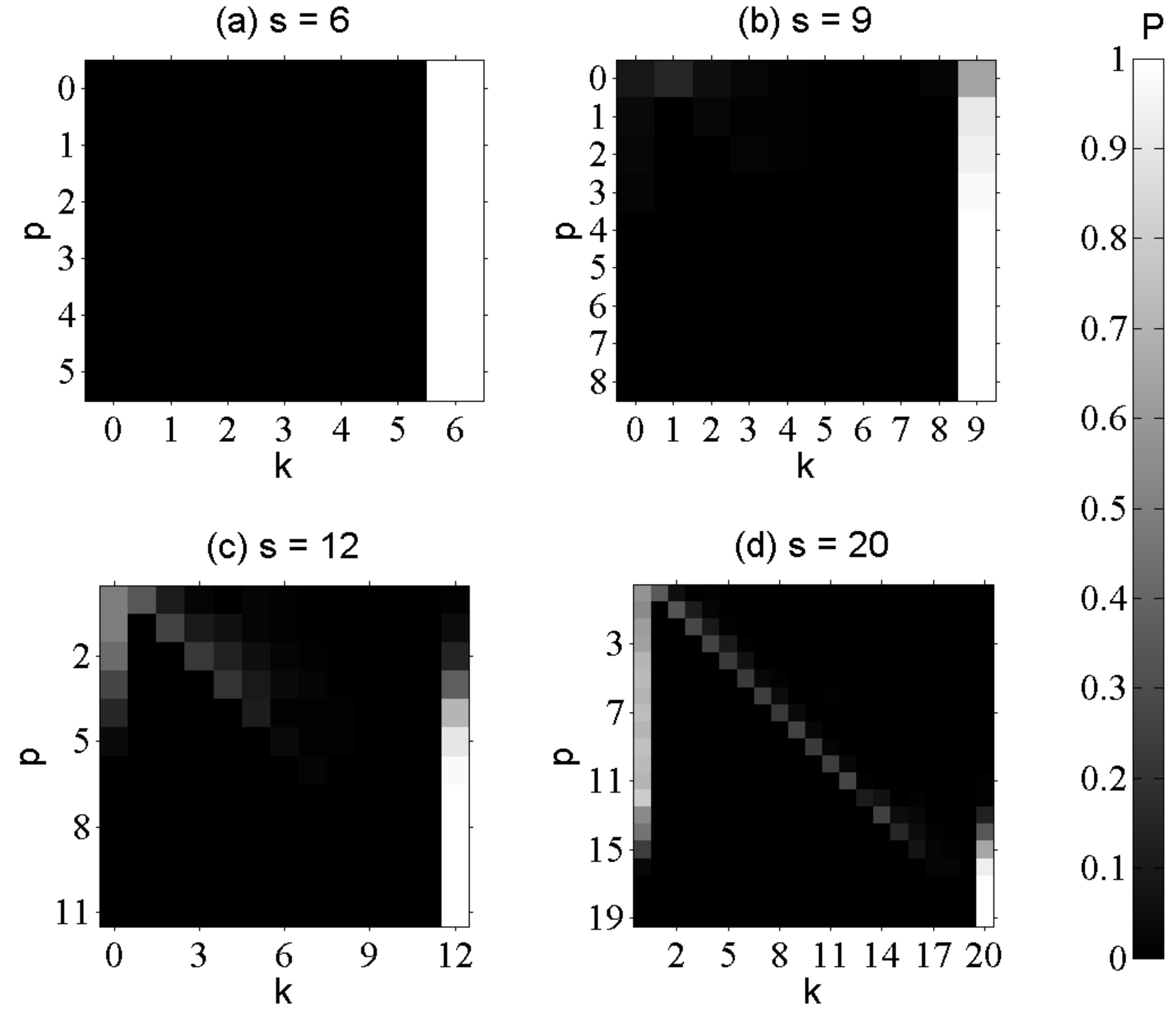}
\caption{The probability $P$ recovering the unknown spectral content for selected values of $s$. The probability is computed for 100 random realizations of the signal with $(n, m) = (256, 40)$. (The lower diagonal cases when $k > 0$, $k \le p$ are invalid, and do not contribute to the result.)}
\label{fig:highdim}
\end{figure}

\begin{figure} \centering
\includegraphics[width=0.45\textwidth]{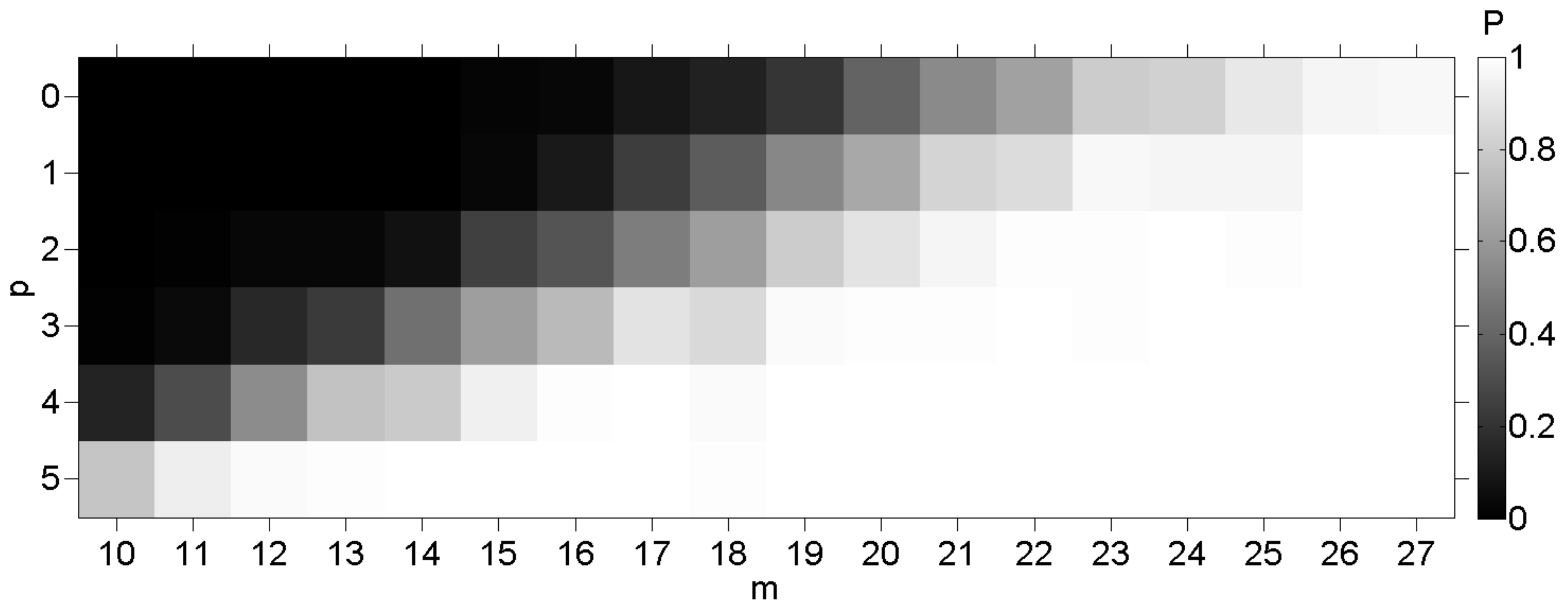}
\caption{A higher probability $P$ of recovering all the unknown frequency content can be achieved with a smaller number $m$ of random observations using the \textit{known-poles} algorithm. The probability is computed for 100 random realizations with $(n, s) = (80, 6)$.}
\label{fig:randomset}
\end{figure}

\begin{figure} \centering
\includegraphics[width=0.45\textwidth]{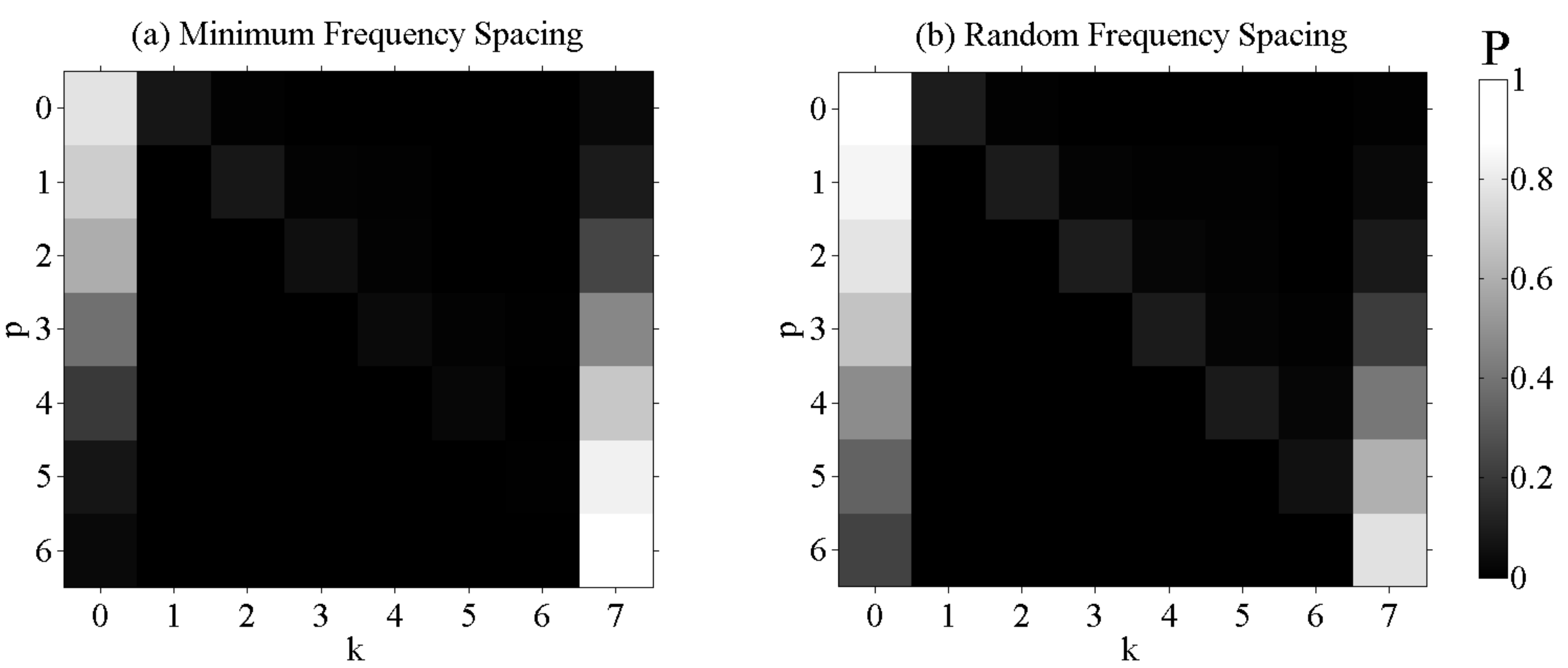}
\caption{Performance of the \textit{known-poles} algorithm when the frequencies do not satisfy any nominal resolution conditions. The probability $P$ of successfully recovering $k$ frequencies is computed for 1000 realizations of the signal with dimensions $(n, m, s) = (40, 15, 7)$. (a) $\Delta f = \nicefrac{1}{(n-1)}$ (b) Frequencies are selected uniformly at random in the band $[0, 1]$.}
\label{fig:freqspacing}
\end{figure}

The phases of the signal frequencies were sampled uniformly at random in $[0, 2\pi)$. The amplitudes $|c_j|, j = 1, \cdots, s$ were drawn randomly from the distribution $0.5 + \chi^2_1$ where $\chi^2_1$ represents the chi-squared distribution with 1 degree of freedom.\\
\textbf{Experiment 1.} We simulated a low-dimensional model with the triple $(n, m, s) = (32, 9, 4)$ and first solved the semidefinite program (\ref{eq:semiotg}) which does not use any prior information, i.e., $p = 0$. For the same realization of the signal, we then successively increase $p$ up to $s-1$, and solve the optimization (\ref{eq:semiprior}) of the \textit{known-poles} algorithm. At every instance of solving an SDP, we record the number $k$ of successfully recovered frequencies along with their complex coefficients. This number also includes the known frequencies if the recovery process returns exact values of their complex coefficients. $k = s$ corresponds to \textit{complete success}, i.e., recovering all of the unknown spectral content. $k = 0$ is \textit{complete failure}, including the case when the complex coefficients of the known frequencies could not be recovered. Figure \ref{fig:lowdim} shows the probability $P$ of recovering $k$ frequencies over $1000$ trials. We observe that even though the complex coefficients of the known frequencies are unknown, the \textit{known-poles} algorithm increases the probability of accurately recovering all or some of the unknown spectral content.\\
\textbf{Experiment 2.} We repeat the first experiment for the higher-dimensional pair $(n, m) = (256, 40)$ and vary $s$. The probability $P$ over 100 random realizations of the signal is shown in Figure \ref{fig:highdim} for selected values of $s$. We observe that the probability of successfully recovering all the frequencies using the \textit{known-poles} Algorithm \ref{alg:freqLocal} increases with $p$.\\
\textbf{Experiment 3.} Figure \ref{fig:randomset} shows the probability $P$ of \textit{complete success}, (i.e. $k = s$), as a function of $m$ over 100 trials for the twin $(n, s) = (80, 6)$. We note that, the \textit{known-poles} algorithm achieves the same recovery probability when compared to (\ref{eq:semiotg}) with a smaller number of random observations.\\
\textbf{Experiment 4.} We now consider these two cases: (a) when $\Delta f = \nicefrac{1}{(n-1)}$, the resolution limit for the numerical experiments in \cite{tang2012csotg}, and (b) when the frequencies are drawn uniformly at random and do not adhere to any minimum resolution conditions. Figure \ref{fig:freqspacing} shows the probability $P$ of recovering $k$ frequencies over 1000 trials for the triple $(n, m, s) = (40, 15, 7)$. We note that the probability of \textit{complete success} with \textit{known poles} suffers relatively little degradation for the random frequency resolutions. These trials include instances when the minimum resolution condition does not hold, formulation in (\ref{eq:semiotg}) shows \textit{complete failure} but the \textit{known-poles} algorithm recovers the unknown spectral content with \textit{complete success}.
\section{Summary}
\label{sec:summary}
Our off-the-grid CS formulation for the \textit{known-poles} algorithm suggests that using the known frequencies of the signal, the signal recovery performance can be improved leading to a higher probability of recovering some or all the residual spectral information - frequency, amplitude and phase - with smaller number of random observations. In the future, it would be interesting to further investigate theoretical performance limits of off-the-grid compressed sensing with prior information.
\section{Acknowledgement}
\label{sec:ack}
The authors would like to thank Gongguo Tang and Benjamin Recht of University of Wisconsin at Madison for helpful e-mail discussions related to their work in \cite{tang2012csotg}.
\bibliographystyle{IEEEbib}
\bibliography{refs}

\end{document}